\RequirePackage{lineno}
\documentclass[aps,twocolumn,showpacs,preprintnumbers,amsmath,amssymb]{revtex4}
\usepackage{graphicx}
\usepackage{dcolumn}
\usepackage{epstopdf}
\usepackage{color}
\usepackage{amssymb}
\usepackage{bm}
\usepackage{subfigure}
\definecolor{dgreen}{cmyk}{1.,0.,1.,0.2}        
\definecolor{orange}{cmyk}{0.,0.353,1.,0.}    

\def\bea {\begin{eqnarray}}
\def\eea {\end{eqnarray}}

\def\be {\begin{equation}}
\def\ee {\end{equation}}

\newcommand{\Npart}{$N_{\rm part}$}
\newcommand{\aveNpart}{$\langle N_{\rm part}\rangle$}
\newcommand{\sNN}{$\sqrtsign s_{\rm NN}$}
\newcommand{\Ss}{$S\sigma$}
\newcommand{\KV}{$\kappa \sigma^2$}

\begin{document}
\title{ Baseline study for higher moments of net--charge
  distributions at RHIC energies }

\author{Nihar R. Sahoo, Sudipan De and Tapan K. Nayak}
\medskip

\affiliation{Variable Energy Cyclotron Center, Kolkata - 700064, India}

\bigskip


\begin{abstract}

Lattice QCD models predict the presence of a critical point in the QCD 
phase diagram where the first order phase transition between the
hadron gas and Quark-Gluon Plasma ceases to exist.
Higher moments of conserved quantities, such as net--charge,
net--baryon number and net--strangeness, are
proposed to be sensitive probes for locating the critical point. The
moments of net--charge distributions have been
studied as a function of centrality for \mbox{Au+Au} collisions at $\sqrt{s_{\rm NN}}$ = 7.7 to 200 GeV 
using three event generators, {\it viz.}, UrQMD, HIJING, and
THERMINATOR-2.  The effect of centrality selection, resonance production, 
as well as contributions from particle species to the net--charge moments and their
products have been studied. It is observed that mean of the
net--charge distributions
are dominated by net--protons, whereas standard
deviation, skewness and kurtosis closely follow net--pion distributions.
These results, along with the predictions from
Hadron Resonance Gas (HRG) model, are presented.

\end{abstract}

\pacs{25.75.-q,25.75.Gz,25.75.Nq,12.38.Mh}
\maketitle

\section{Introduction}

Statistical QCD (Quantum Chromodynamics) calculations predict that at high temperature and/or
energy density, a system of strongly interacting particles, consisting
of quarks and gluons, is formed where the particles would interact
fairly weakly due to asymptotic freedom. Such a phase consisting of (almost) free quarks and
gluons is termed as the Quark Gluon Plasma (QGP).
Experimentally the QGP can be produced by colliding heavy nuclei at
ultra-relativistic energies, thus creating a region of enormous energy density.
Understanding the nature of phase
transition from normal hadronic matter to QGP 
has been a topic of tremendous interest over last decades, both in terms of 
theoretical studies and large scale experiments.
The Relativistic Heavy-Ion Collider (RHIC) at Brookhaven National
Laboratory has been the major facility for the search 
and study of QGP and to explore the QCD phase transition. 
Theoretical models, based on lattice QCD, reveal that
at vanishing $\mu_{B}$, the transition from QGP to hadron gas is a simple 
crossover~\cite{aoki}, whereas at large $\mu_{B}$, the phase
transition is of first order~\cite{aoki,ejiri,bowman,stephanov,fodor,gavai,cheng,indication}.
Therefore, one expects the existence of a critical point at the end of the
first order transition. Locating the QCD critical point has been one of
the major thrusts of the physics program at RHIC. The collider is
capable of accelerating and colliding beams from top energy 
at $\sqrt{s_{\rm NN}}$ = 200~GeV, down to as low
as $\sqrt{s_{\rm NN}}$ = 7.7~GeV.
Taking advantage of this, a beam energy scan program has been undertaken to exploit 
wide region of phase diagram and to search for the possible location
of the QCD critical point~\cite{STAR1,STAR2,qm2012,ismd,STAR_BES}.

One of the most plausible signatures of the critical point has been
predicted to be
the large event-by-event fluctuations of thermodynamic quantities measured
in high energy heavy--ion collisions~\cite{rajagopal}.
This is because of the fact that the thermodynamic
susceptibilities ($\chi$) 
and the correlation length ($\xi$) of the system diverge at the
critical point~\cite{cheng,stephanov,stephanov2009}.
Various QCD inspired models predict that higher moments
of conserved quantities, such as distributions of net--charge,
net--baryon and the net--strangeness 
are associated with the higher order thermodynamic susceptibilities
and exhibit strong dependence on the correlation strength. 
This makes the moments, such as, mean ($M$), standard deviation
($\sigma$), skewness ($S$) and kurtosis ($\kappa$), etc., 
sensitive probes of the location of the critical point.
These moments are related to the correlation length by
 $\sigma^{2} \sim \xi^{2}$, $S \sim \xi^{4.5}$ and $\kappa \sim \xi^{7}$~\cite{cheng,stephanov2009}. 
In order to cancel the volume term in the susceptibilities, different
combinations of the moments are constructed, such as, 
$ S\sigma = \chi^{(3)}/ \chi^{(2)}$ and $\kappa\sigma^{2} =
\chi^{(4)}/ \chi^{(2)}$, which can be used to have direct comparison
of experimental results to lattice calculations.
Thus at the critical point, one would
expect large non-monotonic behavior in these products.
Recent lattice QCD model estimations~\cite{freezout_karsch} have proposed
the extraction of freeze-out parameters of the collision from
the analysis of higher moments of net--charge distribution. 

Among all conserved quantities, experimentally it is advantageous
to study net--charge distributions. 
The fluctuations of net--charge include effects from net--baryon and
net--strangeness. Net-charges take into account all the
charged particles produced in the collisions, and thus can be directly
compared with theoretical calculations~\cite{skokov,kitazawa,Bzdak1}. 
Analytical calculations~\cite{Bzdak} also suggest that the net--charge higher moment
analysis may be less affected by acceptance effects. 
To understand various physics processes which contribute to the
moments of net--charge distributions, three different models have been used,
such as,  a QCD based model (HIJING~\cite{HIJING}), a transport
model (UrQMD~\cite{Bass})
and a thermal model, THERMINATOR-2~\cite{therminator}.
The results of these event generators are compared to those of the 
Hadron Resonance Gas (HRG)~\cite{HRG} model predictions.
Because of the absence of any critical phenomenon in these models,
they set a baseline
for the measurements at the Relativistic Heavy-Ion Collider (RHIC).
Rest of the article is organized as follows. 
Various moments of the net--charge distributions are introduced in
Section II.  
Importance of centrality selection and centrality bin width correction
are presented in Section III. 
In section IV, contributions of different
particle species on the net--charge moments have been discussion, and
in section V, effect of resonance decays are studied.
In section VI, 
the products of moments from different event generators are presented and compared to that of
the HRG model. Finally, we conclude with a summary.

\section{Moments of net--charge distributions} 

Experimentally, both the positive ($N_{+}$) and negative
($N_{-}$) charged particles are measured in a finite acceptance on an
event-by-event basis. Net--charge ($N$) is defined as the 
difference between the number positive and negative charged
particles, $N=N_{+}-N_{-}$. For a given distribution of $N$, the first
four moments can be expressed as:
\begin{eqnarray}
{\rm Mean:~~~~}  M & = & \langle N \rangle,  \nonumber   \\
{\rm variance:~~~~}   \sigma^{2} & = &   \langle (\delta N)^{2} \rangle,   \nonumber   \\
{\rm skewness:~~~~}    S & = & \frac{\langle (\delta N)^{3}\rangle }{  \sigma^{3}}, \nonumber \\
{\rm and,~~~~}  {\rm kurtosis:~~~~}    \kappa & = &  \frac{\langle (\delta
  N)^{4}\rangle}{\sigma^{4} }-3, 	
\end{eqnarray}
where $\delta N \  = \ N-\langle N\rangle $. 
The mean, variance, skewness and kurtosis are measures of the most probable
value, width, asymmetry and the peakedness of the distributions, respectively.


The net--charge distributions in a heavy-ion collision are sensitive
to the collision centrality, which is normally expressed in terms of impact
parameter ($b$) or the number of participating nucleons (\Npart). 
A given centrality class corresponds to certain percentage 
of total geometrical cross section of the collision, and represented
by \aveNpart.

\begin{figure}[tbp]
 \centering
 \includegraphics[width=0.4\textwidth]{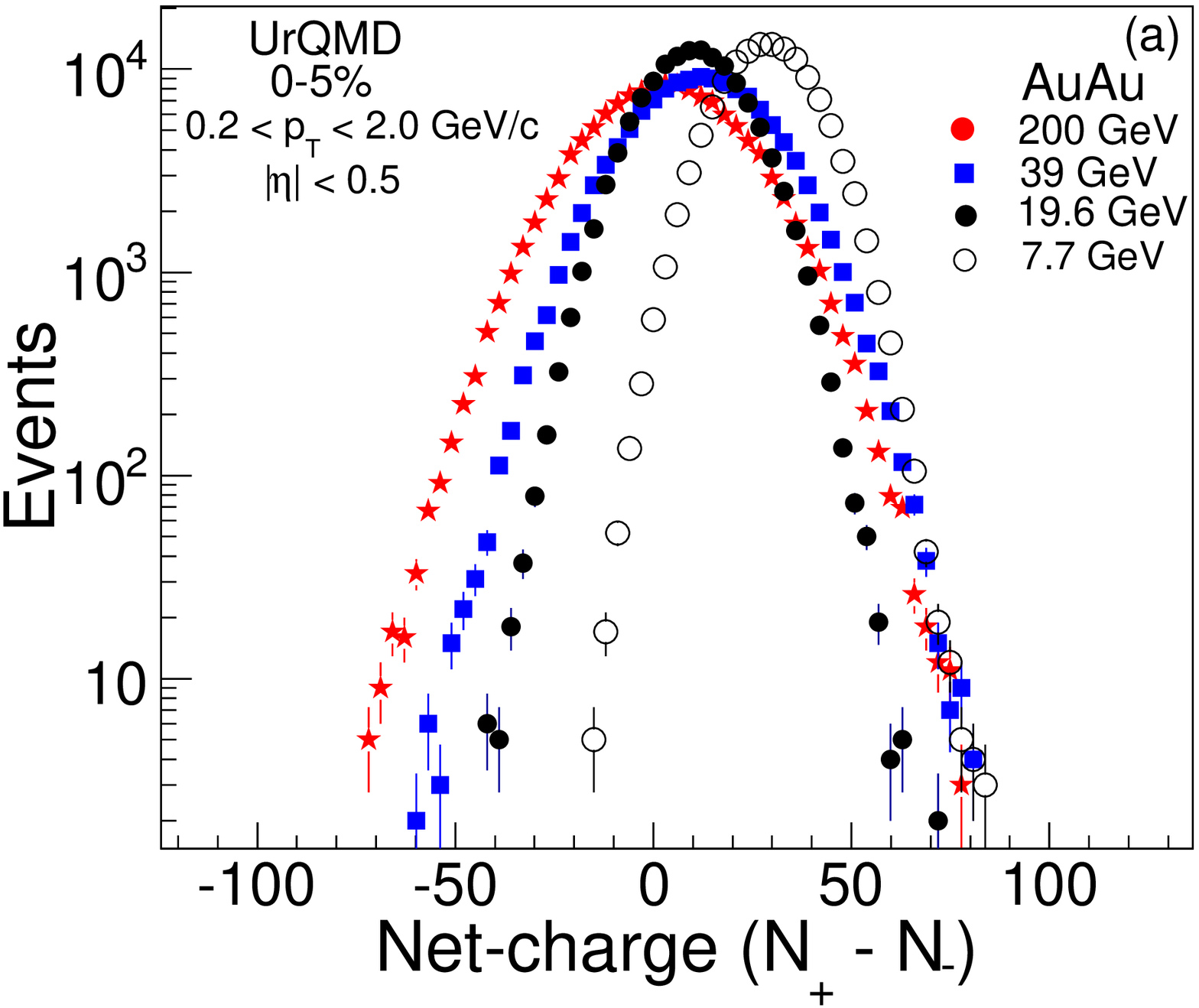} 
 \includegraphics[width=0.4\textwidth]{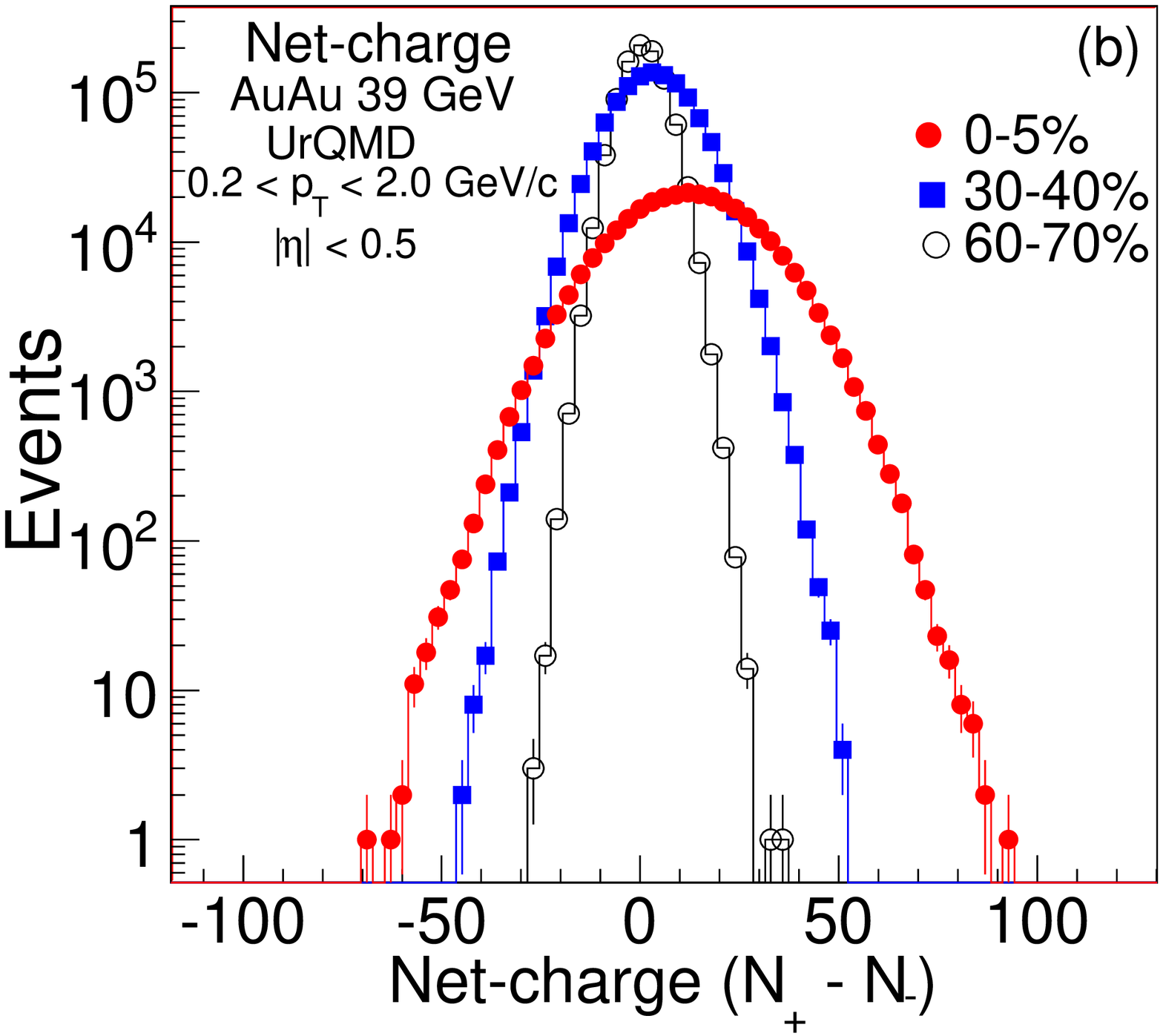} 
\caption{(Color online). The net--charge distributions 
obtained from UrQMD for \mbox{Au+Au} collisions: (a) for central collisions at 
$\sqrt{s_{\rm NN}}$=7.7, 19.6, 39 and 200 GeV, and (b)
for three centralities (0-5\%, 30-40\% and 60-70\% of total cross
section) at $\sqrt{s_{\rm NN}}$=39 GeV.
}
\label{fig1}
\end{figure}

In this study, UrQMD event generator is used to study the 
beam energy and collision centrality dependence of net--charge distributions
for \mbox{Au+Au} collisions. Charged particles within a pseudo-rapidity range 
of $|\eta| < 0.5$ and transverse momentum range of
$0.2 <p_{\rm T}<2.0$~GeV/c are considered. 
Figure~\ref{fig1}(a) shows the net--charge distributions for top
central (0-5\% of total cross section) collisions for four
center-of-mass energies. 
It has been observed that the mean of the distributions is
close to zero for high energy collisions and shifts towards positive values for lower energies. 
The distributions are seen to be wider for higher energy collisions
compared to those of the lower energies. 
Figure~\ref{fig1}(b) shows net--charge distributions for three different
centrality classes for \mbox{Au+Au} collisions at \sNN=39~GeV. 
It is seen that from peripheral to central collisions, the mean as
well as the width of the distributions increase. 

\section{Centrality bin width effect on higher moments}

\begin{figure}[tbp]
 \centering
 \includegraphics[width=0.45\textwidth]{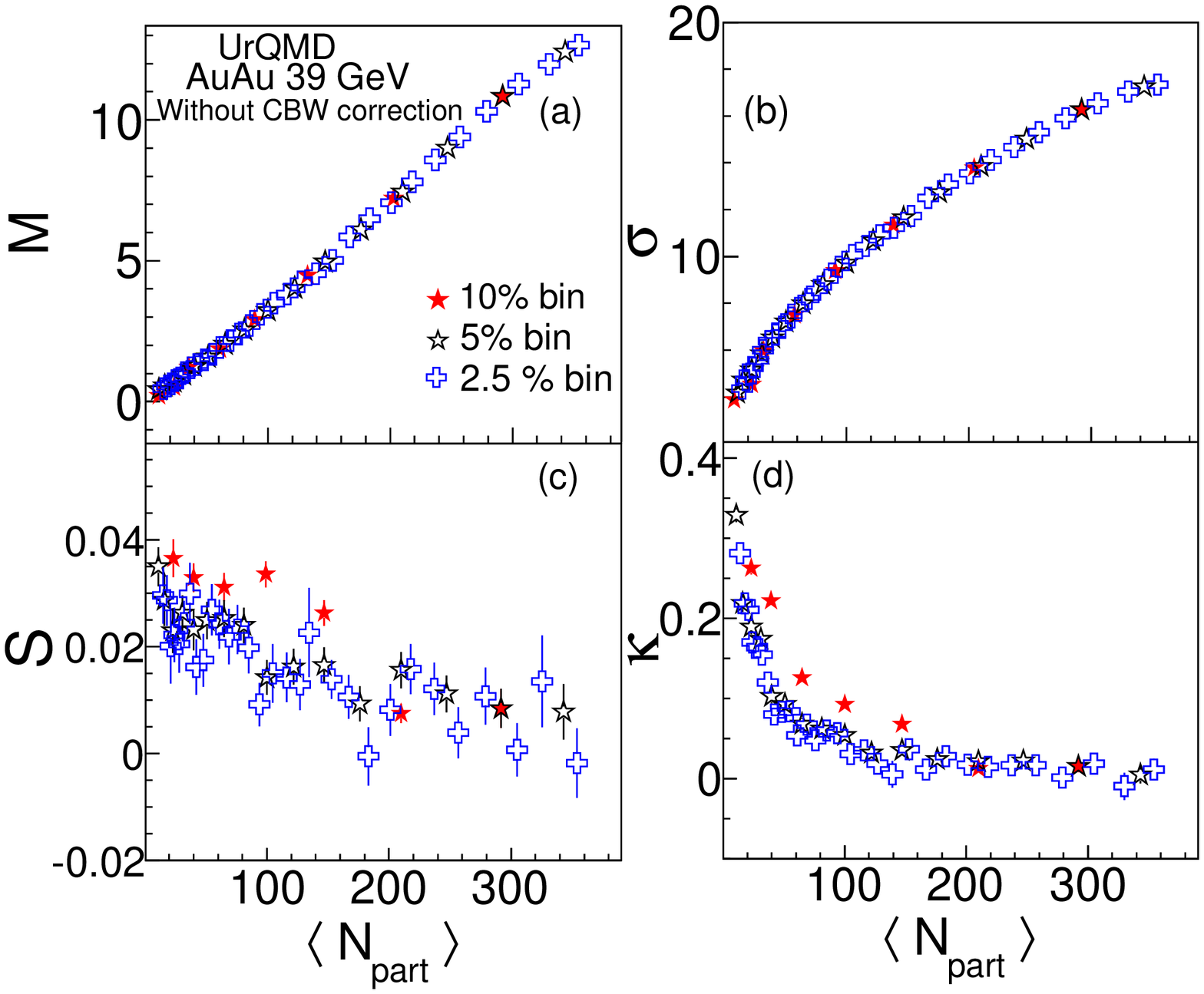}
 \includegraphics[width=0.46\textwidth]{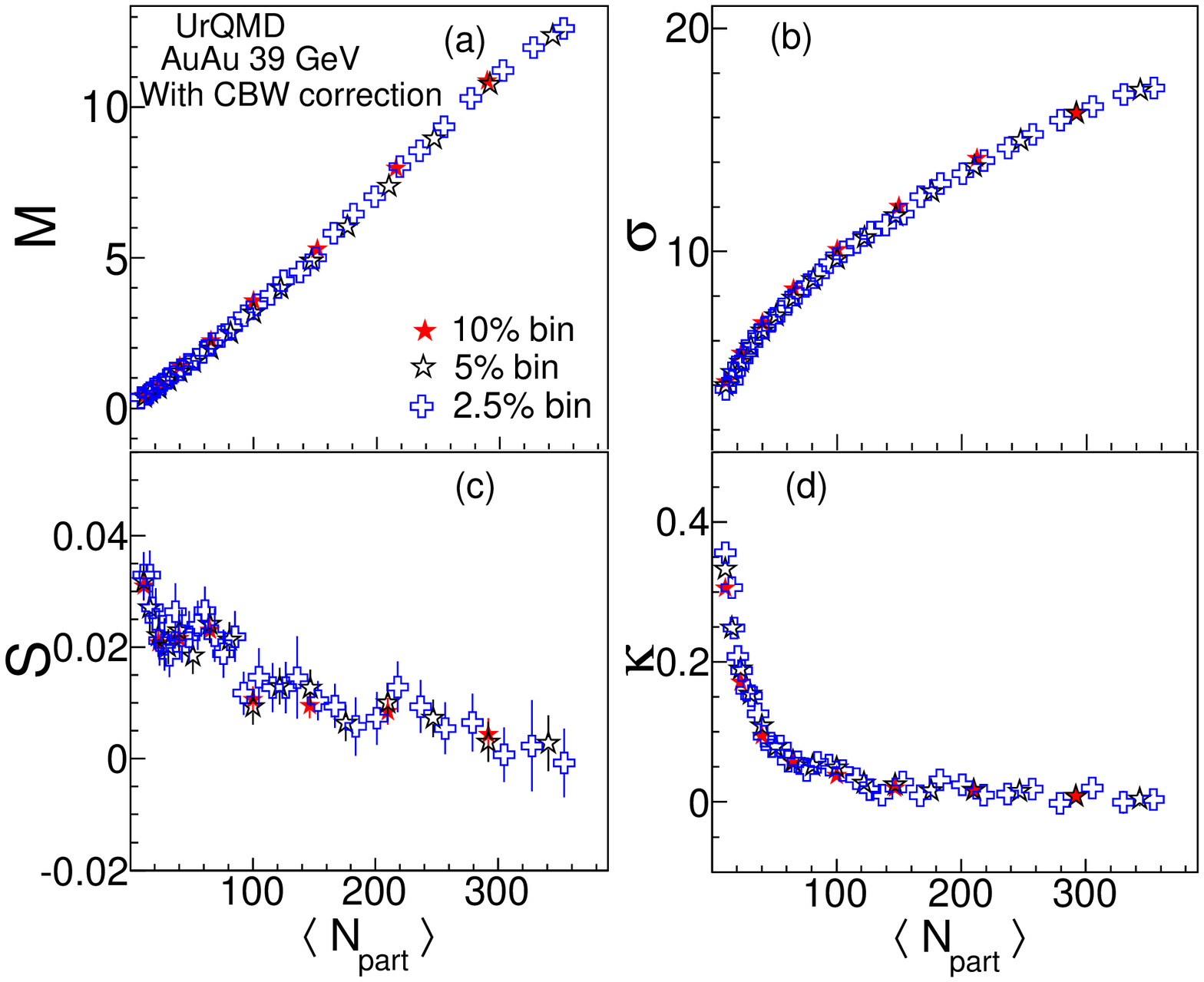} 
 \caption{(Color online). Centrality dependence of (a) mean, (b) standard
deviation, (c) skewness, and (d) kurtosis, obtained for 
\mbox{Au+Au} collisions at $\sqrt{s_{\rm NN}}$=39 GeV.
The results using three centrality bins (2.5\%, 5\% and 10\%), without
and with the centrality bin width corrections are shown in upper and
lower panels, respectively.
}
 \label{fig2}
\end{figure}

A given centrality class is a collection of events having a range of impact parameters or
\Npart, thus comprising of events with different charged particle multiplicities.
This results in additional fluctuations in the number of produced
particles within each centrality class. 
Below we discuss the effect of finite bin width of a given centrality
class on the moments of distributions and prescribe a method in order
to correct for this. 

Figure~\ref{fig2} shows the
centrality dependence of mean, sigma, skewness and
kurtosis of net--charge distributions for \mbox{Au+Au} collisions at
\sNN=39~GeV, obtained using UrQMD. 
Centrality classes were chosen using three different centrality bin
widths, {\it viz.}, 2.5\%, 5\% and 10\% of cross section.
From Fig.~\ref{fig2}(a), it is observed that mean and sigma of the net--charge distributions
are close to each other for all three centrality classes, whereas
deviations are observed for skewness and kurtosis. This deviation is
the result of choosing centrality class with a large bin.

The effect due to the finite centrality bin can be reduced by choosing
smaller bins. In some cases, because of practical reasons such as
resolution of centrality determination, statistics of available events, 
it is not possible to choose fine bins. In this case, a centrality bin
width weighting method~\cite{CBW} can be used to minimize the effect.
For this, net--charge distributions are constructed for finer bins
in centrality within a centrality class. For moments of the
distributions are obtained for each fine bin and 
weighted to get the final moments, according to:
\begin{equation}
  X = \frac{\sum_{i}n_{i}X_{i}}{\sum_{i}n_{i}},
\end{equation}
where $X$ represents a given moment, the index $i$ runs over each fine centrality bin,
$n_{i}$ is the number of events in the $i^{\rm th}$ bin, and
$\sum_{i}n_{i}$ is the total number of events in the given centrality class.
Figure~\ref{fig2}(b) shows the 
moments of net--charge distributions in each centrality class
after recalculating with appropriate weighting using centrality bin width method. 
After this correction, no centrality bin width dependence is
observed in the three centrality classes. Thus the correction method
does an appropriate job in correcting for the finite centrality bin width.
For the rest of the article, results will be presented after taking
into account correction for finite centrality bin width.

\begin{figure}[tbp]
 \includegraphics[width=0.5\textwidth]{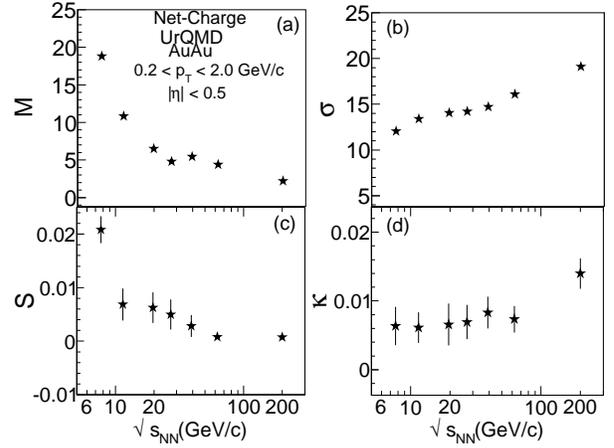} 
 \caption{Beam energy dependence of (a) mean, (b) standard deviation, (c) skewness, and (d)
kurtosis of net--charge distributions, obtained for central (0-5\% centrality bin) \mbox{Au+Au} collisions.
}
\label{species11}
\end{figure}

The beam energy dependence of mean, standard deviation, skewness and
kurtosis of net--charge distributions are presented in
Fig.~\ref{species11} for central (top 5\% of total cross section)
\mbox{Au+Au} collisions. The statistical errors are estimated using the 
Delta theorem~\cite{Luo}.
With the increase of center-of-mass energy from 7.7 to 200 GeV, 
the mean of the distributions decrease, 
whereas standard deviation has a considerable increase.
The skewness decreases with increasing beam energy, whereas kurtosis
shows negligible beam energy dependence.

\section{Effect of particle species on higher moments}

Although we mainly focus on moments of inclusive charged
particles, it is important to obtain the effect of each particle
species which on the total net--charge distributions. 
These species mostly comprise
of $\pi^+$, $\pi^-$, $K^+$, $K^-$, $p$, and $\bar{p}$, so the effect of
net--pion, net--kaon and net--proton distributions on the net--charge
distributions needs to be studied. 
At the generator level, the knowledge of the identity of each
particle makes it possible to perform this study.

\begin{figure}[tbp]
  \includegraphics[width=0.47\textwidth]{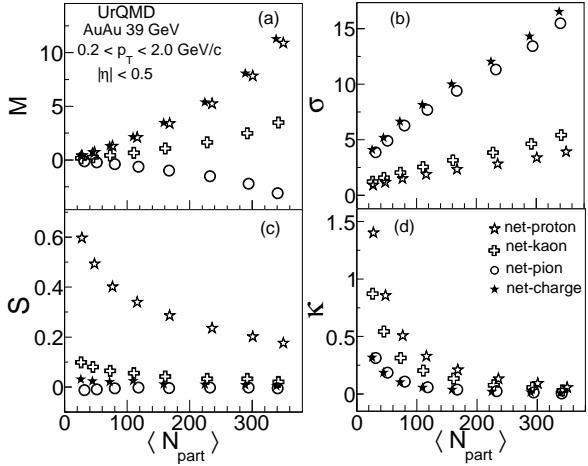} 
  \caption{
Centrality dependence of
(a) mean, (b) standard deviation, (c) skewness, and (d) kurtosis 
for net--charge, net--pion, net--kaon and net-proton distributions
in case of \mbox{Au+Au} collisions at $\sqrt{s_{\rm NN}}$=39 GeV. 
}
\label{species1}
\end{figure}

 \begin{figure}[tbp]
    \includegraphics[width=0.4\textwidth]{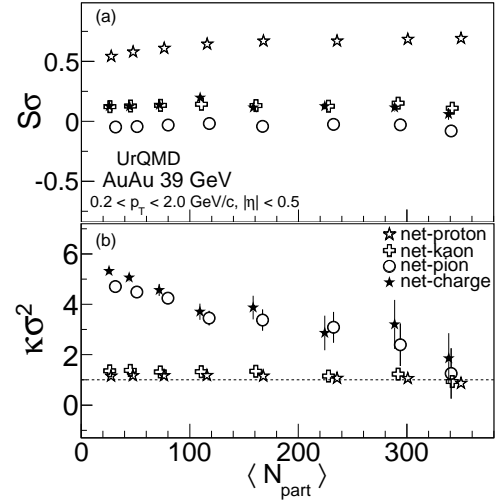} 
  \caption{ Centrality dependence of (a) $S\sigma$ and (b) $\kappa\sigma^{2}$ 
for net--charge, net--pion, net--kaon, and net--proton
distributions in case of \mbox{Au+Au} collisions at $\sqrt{s_{\rm NN}}$=39 GeV.
}
  \label{species2}
\end{figure}

 \begin{figure}[tbp]
   \includegraphics[width=0.43\textwidth]{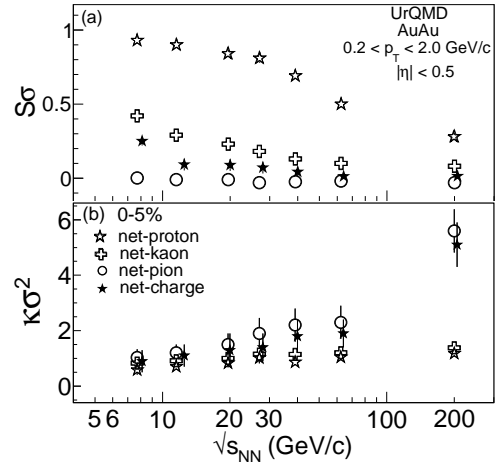} 
   \caption{Beam energy dependence of (a) $S\sigma$ and (b)
     $\kappa\sigma^{2}$ for net--charge, net--pion, net--kaon, net--proton distributions in
    case of central \mbox{Au+Au} collisions.
}
\label{species3}
\end{figure}

The centrality dependence of the contributions from
different particle species to the moments of net--charge distributions have been studied using 
UrQMD for \mbox{Au+Au} collisions at $\sqrt{s_{\rm NN}}$=39~GeV.
Inclusive charged particles as well as identified particles are selected in the 
same pseudo-rapidity ($|\eta| < 0.5$) window and same transverse
momentum ranges ($0.2 <p_{\rm T}<2.0$~GeV/c). 
Figure ~\ref{species1} shows the mean, standard deviation, skewness
and kurtosis for net--charge, net--pion, net--proton and net--kaon distributions
as a function of centrality.  
It is observed that the effect of particle species on the net--charge
distributions are significant, and different species affect the
moments in a different manner.
The mean of the net--charge distributions are dominated by the effect
of net--protons.
The mean of the net--pion distributions, on the other hand, are seen to decrease going
from peripheral to central collisions, whereas the trend is opposite
for the other three cases. The mean for net--pions shift to negative
values for central collisions. The trend for the mean of net-kaons is similar to those of net--charges.
The widths of net--charge distributions are close to those of the
net--pions. The widths of net--kaons and net--protons are close to
each other, but the values are smaller than those of net--charges.
The skewness of the net--charge distributions are
close to those of the net--pions, whereas net--kaon values are not so far.
On the other hand, the 
skewness for net--proton distributions are much larger compared to net--charges and has a
significant centrality dependence.
This may be because of the difference in the number of protons and
anti-protons produced in different centrality classes at this energy.
The kurtosis of net--charge distributions are close to those of
net--pions, and smaller than those of the for net--kaons and net--protons.

As the products of moments (such as \Ss~and \KV) can directly be compared to lattice
calculations, we have studied their centrality dependence in
Fig~\ref{species2} as a function of centrality for \mbox{Au+Au} collisions
at \sNN=39~GeV. It is observed that
the  \Ss~values do not show centrality dependence at this energy.
\Ss~for net--charges are close to those of the net--kaons, whereas  
\Ss~for net--pions are close to zero.
These values for net--protons are much larger compared to net--charges.
The \KV~values for net--charges are close to those of the net-pions,
and show strong centrality dependence.
For net--kaons and net--protons, these values remain close to unity.


The beam energy dependence of  \Ss~and \KV~for top
central \mbox{Au+Au} collisions, obtained using UrQMD, has been presented in 
Fig.~\ref{species3}, where
the upper and lower panels show the results for \Ss~and \KV, respectively.
It is observed that \Ss~for net--pions are close to zero at all
energies, whereas a decreasing trend with increase of energy is
observed for all other cases. For net--charges, \Ss~values get close
to zero at higher energies. The energy dependence of net--proton is quite prominent with much
larger \Ss~values. The values of \KV for net--charge are close to
those of net--pions, and show increasing trend with increase of beam
energy. For net--proton and net--kaon distributions, \KV are close to unity at all energies.

\section{Resonance effects on higher moments }

\begin{figure}[tbp]
\includegraphics[width=0.43\textwidth]{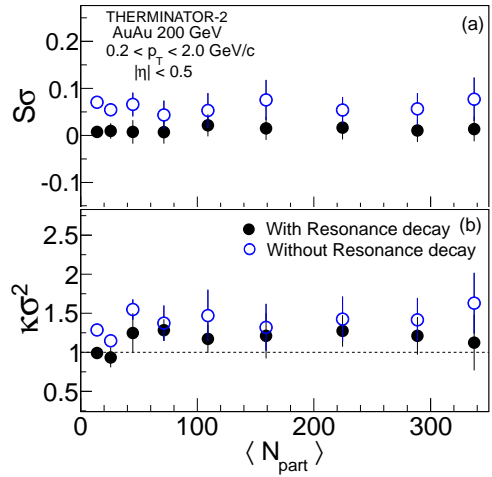} 
\caption{(Color online).
Centrality dependence of (a) $S\sigma$ and (b) $\kappa\sigma^2$ for net--charge distributions, 
with and without resonance decays, obtained using 
THERMINATOR-2~\cite{therminator}.}
\label{resonance}
\end{figure}

The production of charged particles and net--charge distributions in proton-proton and heavy-ion
collisions are affected by resonance decays.  
This can be studied by an event generator where one can track each of
the particles in order to know its origin and history of the decay.
THERMINATOR-2 offers such possibility for \mbox{Au+Au} collisions at
\sNN=200~GeV, where decays of resonances, such as, 
$\Xi$, $\Delta^{++}$, $\rho$, $\phi$ and $\omega$, and their
anti-particles, can be turned on and off.
Net--charge distributions have been made with resonance decays turned
on and off, and the effects on the products of moments are studied.
Figure~\ref{resonance} shows the results for \Ss~and \KV~
as a function centrality.
\Ss~ shows no centrality dependence, and the values 
with resonance decay turned on are closer to zero compared to those without
the decay. Values of 
\KV~without resonance decay are higher compared to those with the decay
at all centralities. The reason for this could be the presence of 
double charged baryons, like  $\Delta^{++}$, which may affect the net--charge
distributions, and enhance the higher order moments~\cite{PBM}.

\section{Comparison of model predictions}

\begin{figure}[tbp]
\includegraphics[width=0.45\textwidth]{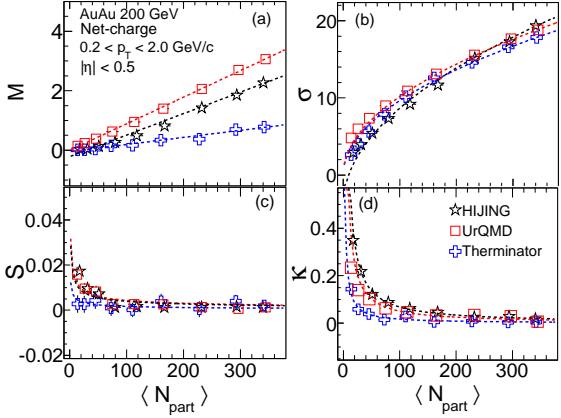} 
\caption{(Color online). 
Centrality dependence of
(a) Mean, (b) standard deviation, (c) skewness and (d) kurtosis 
for net--charge distributions in case of
\mbox{Au+Au} collisions at $\sqrt{s_{\rm NN}}=$200~GeV using 
HIJING,  UrQMD and THERMINATOR-2.
Dashed lines give results using central limit theorem.
}
\label{model1}
\end{figure}

The net-charge distributions are sensitive to the
particle production mechanisms.
HIJING treats the heavy-ion collisions as a superposition of
nucleon-nucleon collisions. 
It is well suited to study the effects of jets and mini-jets on produced
particles.  UrQMD is a hadronic transport model including strings. It has been used
successfully to describe the
stopping power and hadronic re-scattering along with various hadronic resonances.
The Lund string model is used for the particle production both in the HIJING and UrQMD models.
On the other hand, THERMINATOR-2 gives a good description for the
thermal model of particle production. 
The moments and their products for the net-charge 
distributions are analyzed for the three event generators for
\mbox{Au+Au} collisions at
$\sqrt{s_{\rm NN}}$=200~GeV. Figure~\ref{model1} shows the 
mean, standard deviation, skewness and kurtosis as a function of
centrality. Although the trends of all the distributions are similar,
there are some differences in terms of magnitudes. 
The $M$ and $\sigma$ increase from peripheral to
central collisions for all three models, 
whereas $S$ and $\kappa$ values decrease with increase in collision centrality. 
The values of $M$ are lower in THERMINATOR-2 as compared to HIJING and UrQMD,
but the standard deviations are similar in all three cases. The
centrality dependence of $S$ and $\kappa$ for 
THERMINATOR-2 are also weaker as compared to the other two cases.

\begin{figure}[tbp]
\includegraphics[width=0.4\textwidth]{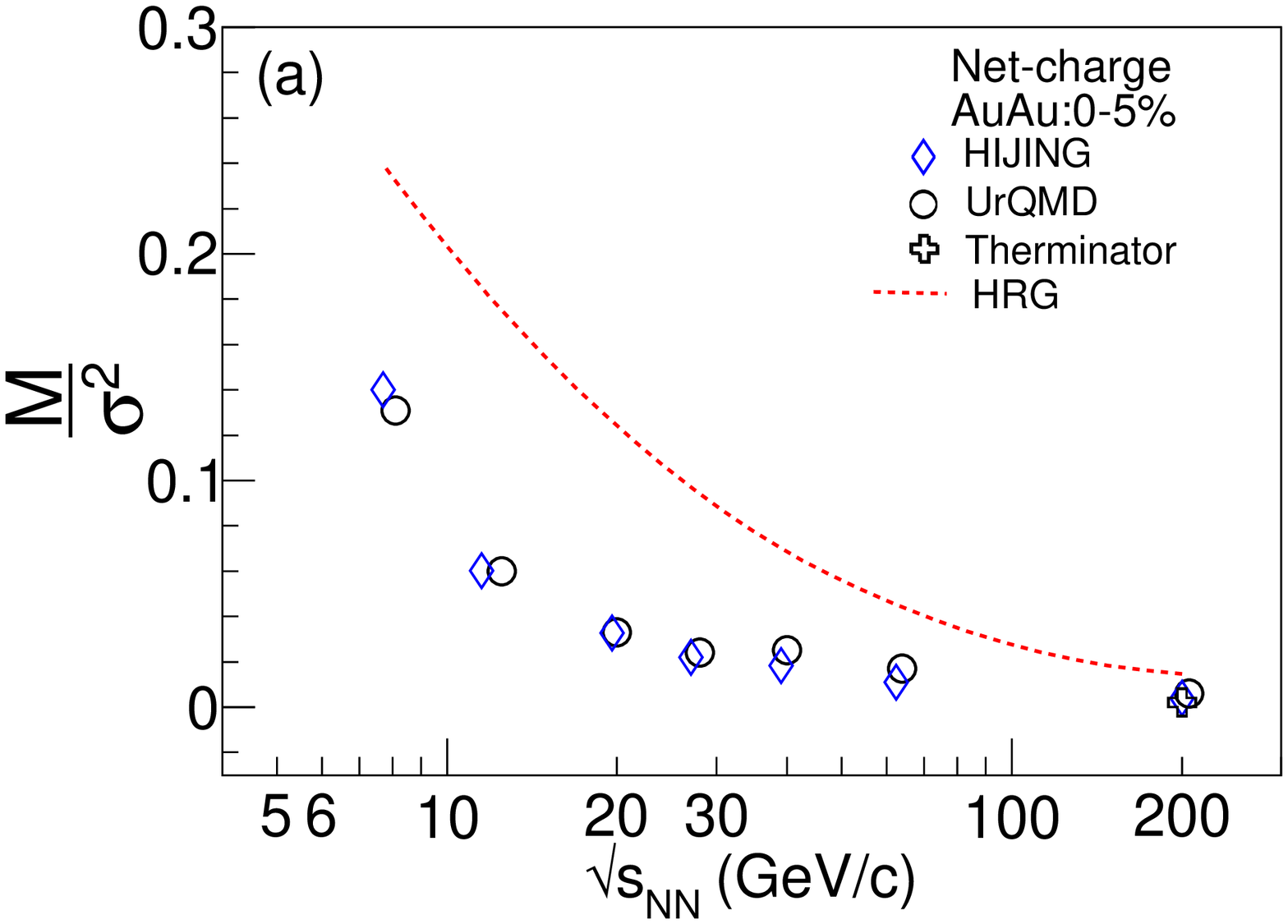} 
\includegraphics[width=0.4\textwidth]{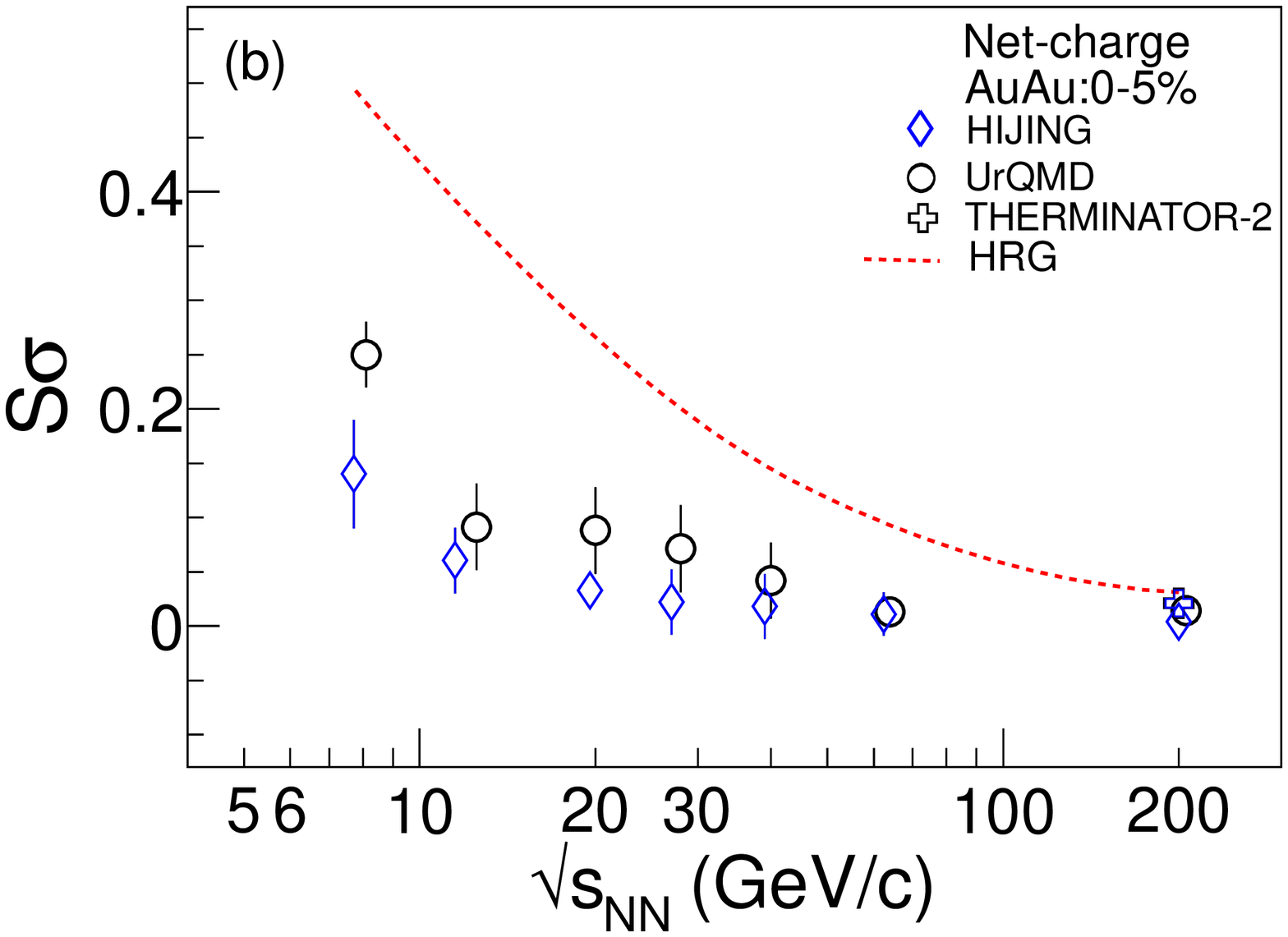} 
\includegraphics[width=0.4\textwidth]{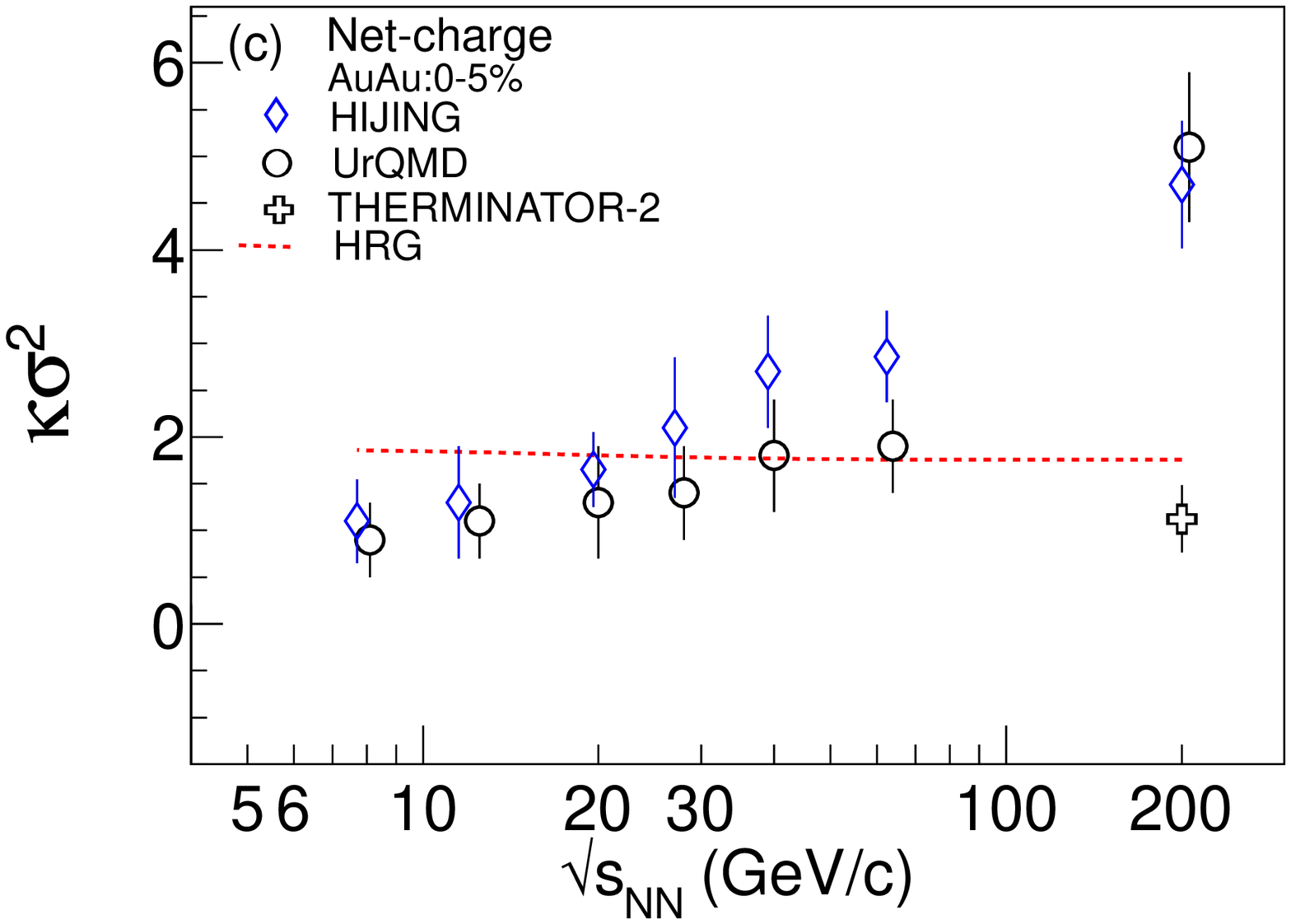} 
\caption{(Color online). 
Beam energy dependence of (a) $M/\sigma^{2}$, (b) $S\sigma$ and 
(c) $\kappa\sigma^{2}$ of
net--charge distributions for central (0-5\%) collisions, obtained from
HIJING, UrQMD, and THERMINATOR-2.
The predictions from the HRG model are superimposed.}
\label{model2}
\end{figure}

The centrality evolution of the higher moments can be better understood by invoking the
Central Limit Theorem (CLT)~\cite{CLT1,CLT2}, which gives the
dependence of the moments on the number of participating nucleons:
\begin{eqnarray}
M \    & \propto & \ \langle N_{\rm part} \rangle,  \nonumber \\
\sigma \ & \propto & \  \sqrt {\langle N_{\rm part} \rangle }, \nonumber \\
S \  & \propto &\ \frac{1}{\sqrt {\langle N_{\rm part} \rangle }},  \nonumber \\
{\rm and} ~\kappa \  & \propto & \ \frac{1}{\langle N_{\rm part} \rangle}.
\end{eqnarray}
The centrality evolution of the moments, as shown by different lines
in Fig.~\ref{model1}, follow the trend of the CLT for all the three event generators. 
 

Combinations of the moments, such as $M/\sigma^{2}$, $S\sigma$ and
$\kappa\sigma^{2}$, have been constructed for central (0-5\% of cross section)
\mbox{Au+Au} collisions at $\sqrt{s_{NN}}$=7.7 to 200 GeV using the three
event generators.
These are shown Fig.~\ref{model2}, along with the predictions from the
HRG model calculations. It is observed that
${M}/{\sigma^{2}}$ and $S\sigma$ decreases with increasing colliding
energy in all cases. The results from the three event generators are close
together and HRG gives higher values.
The values of \KV~from HRG model is seen remain
constant close to 2, whereas variations are seen for the event
generators. The consideration of double charged baryons in HRG model is probably
responsible for \KV~to be close to 2.
HIJING and UrQMD values are close to unity at low energy, after which these
steadily increase as a function of energy. THERMINATOR-2 is available
only at $\sqrt{s_{NN}}$ = 200 GeV, and the value is close to unity.  
At this energy, the models with thermal equilibrium (HRG and
THERMINATOR-2) have produced lower values of \KV compared to 
HIJING and UrQMD.
  
\section{Summary }

Locating the QCD critical point is one of the major tasks in our
understanding of the QCD phase diagram. 
Higher moments of net--charge
distributions are proposed to provide one of the most sensitive
probes towards the search for the critical point. 
The moments and their
products have been studied at RHIC energies for \mbox{Au+Au} collisions
within a range of \sNN=7.7
to 200 GeV using UrQMD, HIJING and THERMINATOR-2 event generators. 
The net--charge distributions are seen to be sensitive to the width of
centrality class, and a centrality bin width correction is needed in
order to take into account inherent fluctuations within the
centrality class. 
The first four moments, {\it viz.}, mean,
sigma, skewness and kurtosis have been determined for all the energies
at regular centrality intervals. 
For a given collision energy, the mean
and sigma of the net-charge distributions increase in going from
peripheral to central collisions, whereas the skewness and kurtosis have decreasing
trends. For top central (0-5\%) collisions, the mean values go down,
whereas sigma values increase  in going
from low to high beam energy. With the increase of energy,
the values skewness have a decreasing trend, and kurtosis values show
negligible beam energy dependence.
Net--charge distributions inherently contain the contributions
from net--pion, net--proton and net--kaon. The effect of these particle
species on the final net--charge moments have been studied using UrQMD.
It is observed that mean of the net--charge is dominated by net--protons
whereas standard deviation, skewness and kurtosis are affected more by
the net-pion distributions. 
The resonance decay contributions to the net-charge  distributions are
studied by the THERMINATOR-2 model. 
The presence of double charged particles like, $\Delta^{++}$, might
enhance the values of \Ss~and \KV.
The various moments by the three 
event generators are compared for \mbox{Au+Au} collisions as a function of centrality
for $\sqrt{s_{\rm NN}}$=200~GeV, and are seen, in general, to follow CLT expectations. 
The products of the moments, $M/\sigma^{2}$, $S\sigma$  and
$\kappa\sigma^{2}$ for 
top (0-5\%) central collisions,
are studied for these three generators and
predictions from the HRG model.
At $\sqrt {s_{NN}}$ = 200 GeV, a clear difference is observed for 
$\kappa\sigma^2$ in case of thermal models like HRG and THERMINATOR-2,
and other non-equilibrium models like HIJING and UrQMD.
While these studies set the baseline for the measurements at RHIC, the
much-awaited experimental data are needed in order to understand the
particle production mechanisms and to probe the QCD critical point.

\medskip

\noindent
{\bf Acknowledgement}

The authors would like to thank Bedangadas Mohanty, Xiaofeng Luo, Lizhu Chen,
Prithwish Tribedy, Nu Xu, Frithjof Karsch, Swagato Mukherjee, Sourendu
Gupta and Rajiv Gavai for their valuable suggestions
during the preparation of the manuscript.  We thank Panos
Christakoglou for careful reading of the manuscript.

\end{document}